\documentclass[12pt]{article}
\usepackage{graphicx}
\usepackage{ifthen}
\usepackage{xstring}
\usepackage{amsmath}
\usepackage{cite}

\newcommand{ \todo}[1]  {{\bf \{TODO~\citen{todo:#1}\}}}
\newcommand{ \comment}[1]{\todo{commentRemoval}}

\newcommand{ \myRef}[1] {\IfBeginWith{#1}{sec}{Sec.~\ref{#1}}{}\IfBeginWith{#1}{fig}{Fig.~\ref{#1}}{}\IfBeginWith{#1}{eq}{Eq.~\eqref{#1}}{}\IfBeginWith{#1}{appx}{Appx.~\ref{#1}}{}}

\newcommand{ \dee}[2]   { \frac{\mathrm{d}{#2}} {\mathrm{d}{#1}} }
\newcommand{ \Dee}[2]   { \frac{\partial{#2}} {\partial{#1}} }
\newcommand{ \OmegaInt}[1]   {\int_{\Omega} {#1}\,\mathrm{d}x}
\def\smallint{\begingroup\textstyle \int\endgroup}
\newcommand{ \A}[2]  { {\smallint_{0}^{x} {#1}\,\mathrm{d} {#2} }}
\newcommand{ \AT}[2]  { {\smallint_{x}^{L} {#1}\,\mathrm{d} {#2} }}
\newcommand { \numberthis} {\addtocounter{equation}{1}\tag{\theequation}}

\newcommand{ \T}{^{\ddagger}}

\DeclareMathOperator{\diag}{diag}
\DeclareMathOperator{\Diag}{Diag}

\numberwithin{equation}{section}
\numberwithin{figure}{section}

\graphicspath{{../resultPlots/}{../resultPlots/dataSet_a/}}

\begin{document}


\title{Resolving Histogram Binning Dilemmas with \\ Binless and Binfull Algorithms}

\author{Abram Krislock$^1$ \\ Nathan Krislock$^2$}

\maketitle

{$^{1}$~Department of Physics,
Stockholm University -- Oskar Klein Centre,
AlbaNova, SE-106 91, Stockholm, Sweden}

{$^{2}$~Department of Mathematical Sciences,
Northern Illinois University,
DeKalb, Illinois 60115, USA
}

\begin{abstract}
The histogram is an analysis tool in widespread use within many sciences, with high energy physics as a prime example. However, there exists an inherent bias in the choice of binning for the histogram, with different choices potentially leading to different interpretations. This paper aims to eliminate this bias using two ``debinning'' algorithms. Both algorithms generate an observed cumulative distribution function from the data, and use it to construct a representation of the underlying probability distribution function. The strengths and weaknesses of these two algorithms are compared and contrasted. The applicability and future prospects of these algorithms is also discussed.
\end{abstract}

\section{Introduction}
\label{sec:Intro}

High energy physics (HEP) research makes common use use of the histogram as a data analysis tool\footnote{Many other areas of scientific research involving statistical analysis also commonly use the histogram. The techniques described in this paper apply to any such similar analysis. For concreteness, in this paper we will restrict our discussion to that of HEP research.}. A great deal of particle physics data, both experimental\cite{PDG:Review2012} and phenomenological\cite{HEPPheno}, is analyzed with histograms.

The histogram is a way of representing the data in such a way as to have it look like and be comparable to the \emph{underlying} probability distribution function (UPDF) of the particle physics reality or model prediction. We will refer to the histogram, or any other representation, of a given data set as an \emph{observed} probability distribution function (OPDF). The UPDF depends on some set of parameters, and so a set of bins for the OPDF histogram are chosen in terms of those parameters.

However, the choice of these bins involves an inherent bias~\cite{Berg}. The analysis of histogrammed data can be highly dependent upon the set of bins chosen. This is particularly true for the case of performing a fit of a histogram with a theoretical or parameterized UPDF; the results obtained from different choices of bin sets may differ more than the reported uncertainty in the fit. 

\begin{figure}[t!]
  \centerline{\includegraphics[width=0.8\textwidth]{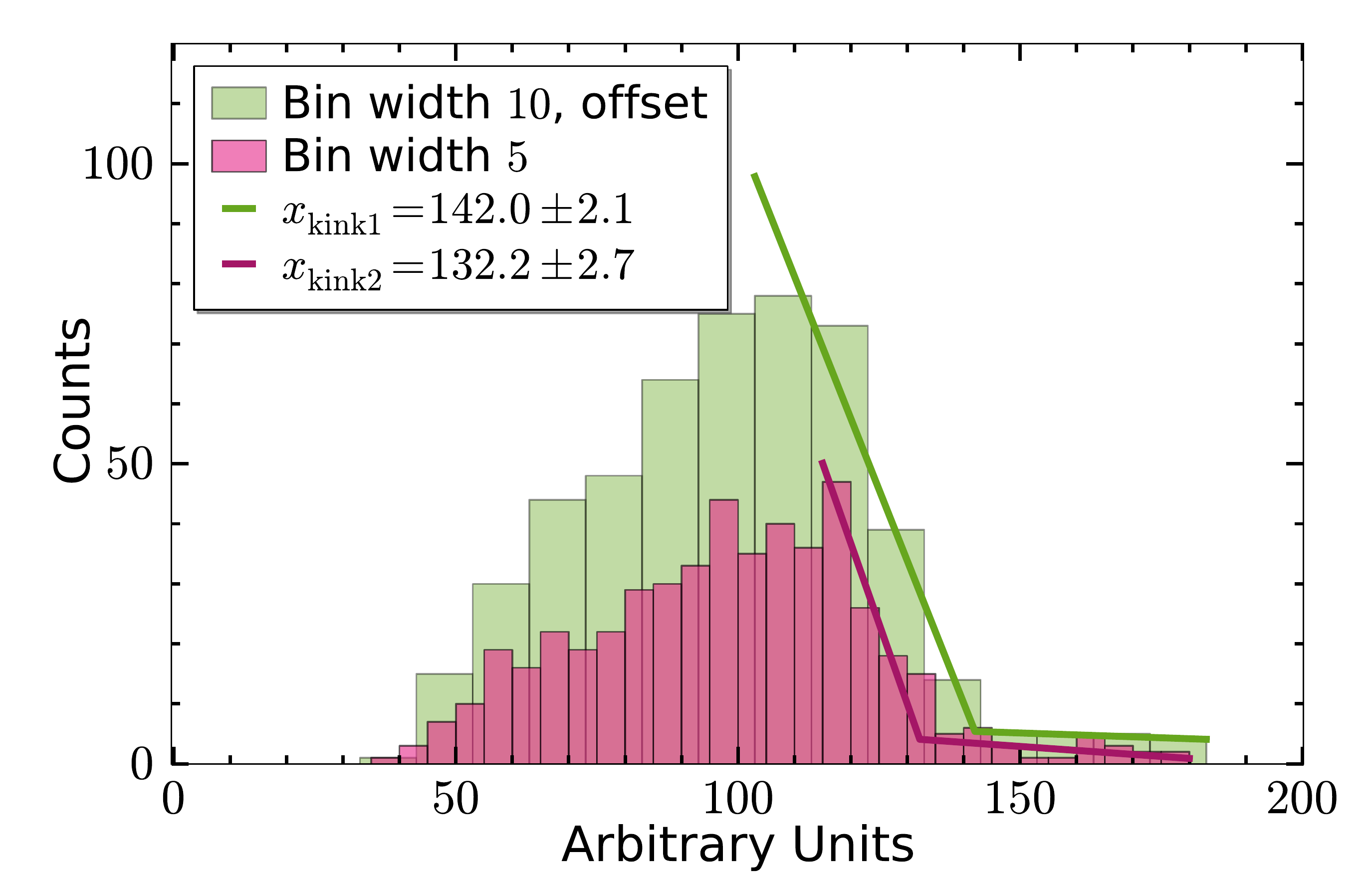}}
  \caption{The same data is used to fill two histograms with different bin sets. \protect\myRef{eq:intersectingLines} is used to fit each of these from their peak bin to their last non-zero bin. The green (light gray) histogram has offset bins, with the first bin edge beginning at \ensuremath{x=3} and a bin width of 10. The purple (gray) histogram has a bin width of 5, beginning at the origin. The fit results for these two histograms are shown as green (gray) and purple (dark gray) lines, respectively. The locations of the kinks, which are shown in the figure legend, disagree at the $99.8\%$ level.}
  \label{fig:binningDilemma}
\end{figure}

We demonstrate this problem with a simple example from HEP phenomenology: Early searches for supersymmetry at the Large Hadron Collider were expecting to find kinks or endpoints within kinematical distributions such as invariant masses\cite{Hinchliffe:SUSYatLHC}, or the $m_{T2}$ variable\cite{Lester:MT2origin, Cho:MT2kink}. We show such a possible signal in \myRef{fig:binningDilemma}, which shows two histograms of the same data but with different bin sets. The two histograms are then fit with the following function:
\begin{equation}
\label{eq:intersectingLines}
  y = \left\{ \begin{array}{cc}
      m_1 (x - x_{\rm kink}) + b, & {\rm if}\ x < x_{\rm kink}, \\
      m_2 (x - x_{\rm kink}) + b, & {\rm if}\ x \ge x_{\rm kink}.
    \end{array} \right.
\end{equation}
This equation describes a bent line with a kink at $x = x_{\rm kink}$. The slopes of the lines on either side of the kink are $m_1$ and $m_2$, and $b$ is the value of the function at the kink. The location of the kink is the only parameter of interest. \myRef{fig:binningDilemma} shows the two histograms fit with this function, as well as the fit result and uncertainty for the parameter $x_{\rm kink}$. As the figure shows, the locations of $x_{\rm kink}$ from the two fits do not agree. If we treat the fit results from the figure as normal distributions, the probability for a sample $x_{\rm kink2}$ to be greater than a sample $x_{\rm kink1}$ is $0.2\%$. Thus, they disagree at the $99.8\%$ level.

Thus, the choice of bin set can certainly affect the outcome of an analysis. Bin sets are typically chosen by eye to be the smallest width bins such that there are ``enough'' statistics in the bins of greatest interest. Ofttimes the bin set is chosen under the constraint of aesthetic rather than scientific reasons. Authors choose uniform bin sizes, usually using bin widths of two, five, or ten times their measurement unit. The bin edges are also chosen to line up with notable x-axis landmarks, such as the origin. The consequences of these choices on scientific results is generally not discussed.

Thus, our goal for this paper is to avoid such bias by constructing OPDF representations other than regular histograms. Our new representations must avoid the above mentioned binning dilemmas. In this paper, we will present two such ``debinning'' algorithms, both of which involve the relations between the OPDF and the observed Cumulative Distribution Function (OCDF). These algorithms were inspired by the method used in \cite{Berg}. The ``binless'' algorithm never bins the data at all, instead determining the OPDF as the smoothed, numerical derivative of the OCDF. In contrast, the ``binfull'' algorithm uses the OCDF as a Monte Carlo generator for the OPDF. A smoothing function is applied during the generation of a very large number of points. These points are then used to create a histogram which is full of very small bins. Each of these two methods has its strengths and weaknesses, which we will also describe.

Currently, both of these algorithms assume an intimate understanding of the backgrounds inherent in a given HEP study. Our algorithms both consider the comparison between the debinned background-only OPDF and the background UPDF. We run each algorithm multiple times over the background data, scanning to find the best-fit value of the smoothing parameter. We then use the resulting smoothing parameter to generate the ultimate debinned full-data OPDF.

The remainder of this paper is organized as follows. In \myRef{sec:binless}, we describe the binless algorithm by deriving its equations, describing their implementation, and presenting example plots. The binfull algorithm is described in \myRef{sec:binfull}, where we describe the Monte Carlo generator, smoothing function, and implementation, as well as presenting more example plots. We conclude, compare these two methods, and comment on future work in \myRef{sec:Conclusion}. Lastly, we have included \myRef{appx:FakeData} to describe the generation of the data used for testing these debinning algorithms. Additional plots, animations, and all of our code can be found on our debinning webpage~\cite{debinning}.

\section{The Binless Algorithm}
\label{sec:binless}

As stated in the introduction, the binless algorithm first forms the OCDF from the data. This is followed by taking the smoothed numerical derivative of it to obtain the binless OPDF.

The method of constructing an OCDF is as follows~\cite{Berg}. First, the data is sorted by value into an array. Thus we have, say, $N$ data values, $x_1, \cdots, x_N$, where $x_i > x_j$ if $i > j$. Then, the OCDF, $z(x)$ is constructed from this data simply as,
\begin{equation}
\label{eq:OCDF}
  z(x) = \left\{ \begin{array}{cl}
      0, & {\rm if}\ x < x_1, \\
      i, & {\rm if}\ x_i \le x < x_{i+1},\ \text{for $i\in[1, N-1]$}, \\
      N, & {\rm if}\ x_N \le x.
    \end{array} \right.
\end{equation}
This function $z$ thus looks like a staircase with stairs of non-uniform length.

Next, we take the derivative of $z$ to arrive at our binless OPDF. The reason it must be a smoothed numerical derivative is that taking a naive numerical derivative of $z$ results in a large amount of noise in the resulting binless OPDF~\cite{Chartrand}. Thus, to obtain a binless OPDF which is even remotely useful, we must have a more sophisticated numerical differentiation algorithm. This algorithm must be able to smooth the noise introduced by numerical differentiation.

\subsection{Numerical Differentiation with Noise Smoothing}
\label{sec:binlessChartrand}

We use the method of numerical differentiation presented in~\cite{Chartrand}, which is a least squares minimization problem including a Total Variation (TV) penalization term~\cite{VogelOman}. The problem is defined as follows: We are working in $\Re$ over some domain $\Omega = [0, L]$, using the inner-product $\langle u, v \rangle = \OmegaInt{ u(x)v(x) }$. We are trying to find the function, $u(x)$, which describes the derivative of some data, $z(x)$, while ensuring that $u(x)$ does not vary too much due to the noise in $z(x)$. The plan is to minimize the following functional\footnote{In this article, whenever its meaning is unambiguous, an apostrophe or ``prime'' in any equation will be understood to be a regular derivative, such that $u'(x) \equiv \dee{x}{u}$.}:
\begin{equation}
\label{eq:ChartrandLeastSquares}
  f[u] = \frac{1}{2} \|Au - z\|_2^{2} + \alpha \|u'\|_1.
\end{equation}
Note that the 2-norm is defined to be $\|v\|_2 = \sqrt{\langle v, v\rangle} = \sqrt{\OmegaInt{ [v(x)]^2 }}$ and the 1-norm is defined as $\|v\|_1 = \OmegaInt{ |v(x)| }$.

The functional in \myRef{eq:ChartrandLeastSquares} contains two terms. The first term, the least squares term, penalizes the difference between the anti-derivative of $u(x)$, where $[A u](x) \equiv \A{u(w)}{w}$, and $z(x)$. The second term is the TV term which minimizes the total amount of variation of $u(x)$. The parameter $\alpha$ controls the relative influence of the TV term compared to the least squares term.

In order to find the $u$ that minimizes the functional $f[u]$ in \myRef{eq:ChartrandLeastSquares}, we first need to determine the  gradient, or functional derivative, of $f$. However, since the TV term in $f[u]$ is not differentiable when $\|u'\|_1 = 0$, we will instead minimize
\begin{equation}
\label{eq:ChartrandExpanded}
  f_\beta[u] = \frac{1}{2} \|Au - z\|_2^{2} + \alpha \OmegaInt{\sqrt{{\left(u'(x)\right)^{2} + \beta^{2}}}},
\end{equation}
where $\beta$ is a small positive parameter.

We remind the reader that the functional derivative can be defined as follows:
\begin{equation}
\label{eq:FunctionalDerivative}
  \Dee{u}{f[u]}(x) = \lim_{h\to0} \frac{f[u + h \delta(\cdot - x)] - f[u]}{h}.
\end{equation}
Here, $\delta$ is the Dirac delta function~\cite{DiracDelta}. For our purposes, the relevant properties of the Dirac delta function are
\begin{align*}
  \OmegaInt{ v(x) \delta(x - y)} &= v(y) \\
  \OmegaInt{ v(x) \delta'(x - y)} &= -v'(y) \\
  \OmegaInt{ v(x)(\A{\delta(w - y)}{w})} &= \int_{y}^{L} {v(x)}\ \mathrm{d}x
  \numberthis \label{eq:DiracDeltaProperty}
\end{align*}
for any function $v$ which is defined over $\Omega$.


Computing the gradient of \myRef{eq:ChartrandExpanded}, using \myRef{eq:DiracDeltaProperty}, is left as an exercise for the reader. The result is
\begin{equation}
\label{eq:ChartrandGradient}
  g[u] \equiv \Dee{u}{f_\beta[u]} = \left[A\T A + \alpha L(u) \right]u - A\T z,
\end{equation}
where the $L$ operator is given by
\begin{equation}
\label{eq:VOL}
  L(u) v \equiv -\left(\frac{v'}{\sqrt{(u')^2 + \beta^2}} \right)',
\end{equation}
and $A\T$ is the adjoint of the $A$ anti-derivative operator, and is given by $[A\T v](x) \equiv \AT{v(w)}{w}$. Note that $A\T$ satisfies $\langle Au, v \rangle = \langle u, A\T v \rangle$.

We define our algorithm by recognizing that the minimum of the least squares functional is found by setting the gradient to zero, and iterating the resulting equation, which is
\begin{equation}
\label{eq:ChartrandIteration}
  \left[A\T A + \alpha L(u^{(k)})\right] u^{(k+1)} = A\T z.
\end{equation}
In this equation, $u^{(k)}$ is the result of the $k$th iteration, starting from some initial guess, $u^{(0)}$. We can define the step of our algorithm to be $s^{(k)} \equiv u^{(k+1)} - u^{(k)}$. Then the algorithm equation becomes
\begin{equation}
\label{eq:Algorithm}
  H^{(k)} s^{(k)} = -g[u^{(k)}], \qquad u^{(k+1)} = u^{(k)} + s^{(k)},
\end{equation}
where $H^{(k)} \equiv \left[A\T A + \alpha L(u^{(k)})\right]$. Since $H^{(k)}$ is the Hessian, or second derivative, of $f_\beta$, this is Newton's method for minimizing $f_\beta$.

\subsection{Implementation for a General Derivative}
\label{sec:binlessImplementation}

Unlike~\cite{Chartrand}, our $x$ values are not uniformly spaced for our data $z(x)$. Thus, some care has to be taken in defining the various operators which go into the algorithm equation, \myRef{eq:Algorithm}. Since our data $z(x)$ is a ``staircase'' function, we treat it as a column vector, $z$, associated with a column vector, $x$. Both of these vectors have length $N$, the number of data points. The operators we need are then defined as square matrices which act on these vectors.

The derivative matrix is based upon the discretization of a simple derivative:
\begin{equation}
\label{eq:SimpleDerivative}
  \dee{x}{z}(x) = \lim_{h\to0} \frac{z(x) - z(x - h)}{h} \ \ 
    \Rightarrow\ \ \left[\dee{x}{z}\right]_{\rm i} 
      \approx \frac{z_{\rm i} - z_{\rm i-1}}{x_{\rm i} - x_{\rm i-1}}.
\end{equation}
The derivative is then defined in terms of a forward difference matrix, $D$, where
\begin{equation}
\label{eq:ForwardDifference}
  D = \left( {\footnotesize
    \begin{array}{cccccc}
      1 & 0 & 0 & \cdots & 0 & -1 \\
      -1 & 1 & 0 & \cdots & 0 & 0 \\
      0 & -1 & 1 & \cdots & 0 & 0 \\
      \vdots & \vdots & \vdots & \ddots & \vdots & \vdots \\
      0 & 0 & 0 & \cdots & 1 & 0 \\
      0 & 0 & 0 & \cdots & -1 & 1
    \end{array}
  } \right).
\end{equation}
(The $-1$ in the upper right corner serves to define a derivative with periodic boundary conditions.) The derivative matrix, $\nabla_x$, is then given as
\begin{equation}
\label{eq:DerivativeMatrix}
  \nabla_x = (\Delta_x)^{-1} D,
\end{equation}
where $\Delta_x$ is a diagonal matrix, with $i$th entry $[\Delta_x]_{\rm ii} = x_{\rm i} - x_{\rm i-1}$ (where $x_{0} \equiv 0$), and $(\Delta_x)^{-1}$ is also diagonal, with $i$th entry $[(\Delta_x)^{-1}]_{\rm ii} = ([\Delta_x]_{\rm ii})^{-1}$.

The $A$ and $A\T$ operators are similarly based upon the discretization of a simple integral. We discretize our integral using trapezoidal areas:
\begin{equation}
\label{eq:SimpleIntegral}
  \left[Az\right]_{\rm i} 
      = \int_0^{x_i} z(w)\,\mathrm{d}w \approx \sum_{j=1}^{i} \frac{1}{2} (x_{\rm j} - x_{\rm j-1}) (z_{\rm j} + z_{\rm j-1}).
\end{equation}
Here, it is again understood that $x_{0} \equiv 0$ and $z_{0} \equiv 0$. Thus, $A$ is defined as
\begin{equation}
\label{eq:A}
  A = \frac{1}{2} \left( {\footnotesize
    \begin{array}{cccccc}
      1 & 0 & 0 & \cdots & 0 & 0 \\
      1 & 1 & 0 & \cdots & 0 & 0 \\
      1 & 1 & 1 & \cdots & 0 & 0 \\
      \vdots & \vdots & \vdots & \ddots & \vdots & \vdots \\
      1 & 1 & 1 & \cdots & 1 & 0 \\
      1 & 1 & 1 & \cdots & 1 & 1
    \end{array}
  } \right) \Delta_x \left( {\footnotesize
    \begin{array}{cccccc}
      1 & 0 & 0 & \cdots & 0 & 0 \\
      1 & 1 & 0 & \cdots & 0 & 0 \\
      0 & 1 & 1 & \cdots & 0 & 0 \\
      \vdots & \vdots & \vdots & \ddots & \vdots & \vdots \\
      0 & 0 & 0 & \cdots & 1 & 0 \\
      0 & 0 & 0 & \cdots & 1 & 1
    \end{array}
  } \right).
\end{equation}
Since
\begin{equation}
\label{eq:SimpleAdjointIntegral}
  \left[A\T z\right]_{\rm i} 
      = \int_{x_i}^{L} z(w)\,\mathrm{d}w \approx \sum_{j=i+1}^{L} \frac{1}{2} (x_{\rm j} - x_{\rm j-1}) (z_{\rm j} + z_{\rm j-1}),
\end{equation}
we find that $A\T$ is similarly defined as
\begin{equation}
\label{eq:AT}
  A\T = \frac{1}{2} \left( {\footnotesize
    \begin{array}{cccccc}
      0 & 1 & 1 & \cdots & 1 & 1 \\
      0 & 0 & 1 & \cdots & 1 & 1 \\
      0 & 0 & 0 & \cdots & 1 & 1 \\
      \vdots & \vdots & \vdots & \ddots & \vdots & \vdots \\
      0 & 0 & 0 & \cdots & 0 & 1 \\
      0 & 0 & 0 & \cdots & 0 & 0
    \end{array}
  } \right) \Delta_x \left( {\footnotesize
    \begin{array}{cccccc}
      1 & 0 & 0 & \cdots & 0 & 0 \\
      1 & 1 & 0 & \cdots & 0 & 0 \\
      0 & 1 & 1 & \cdots & 0 & 0 \\
      \vdots & \vdots & \vdots & \ddots & \vdots & \vdots \\
      0 & 0 & 0 & \cdots & 1 & 0 \\
      0 & 0 & 0 & \cdots & 1 & 1
    \end{array}
  } \right).
\end{equation}

Finally, care also must be taken in constructing the $L(u^{(k)})$ operator. Since it depends on the iteration guess $u^{(k)}$, it must be re-initialized for each step of the algorithm. Considering \myRef{eq:VOL}, we see that derivatives appear in both the numerator and denominator within $L$. We can approximate this ``derivative ratio'' with a ``ratio of forward difference matrix operations''. Thus, the discretization of $L$ is
\begin{equation}
\label{eq:Lu}
  L(u^{(k)}) = - \nabla_x \left(\widetilde{\Delta}_{u^{(k)}}\right)^{-1} D
\end{equation}
where $\widetilde{\Delta}_{u^{(k)}}$ is a diagonal matrix, such that $\left[\widetilde{\Delta}_{u^{(k)}}\right]_{\rm ii} = \sqrt{[Du^{(k)}]_{\rm i}^2 + \beta^2}$.

\subsection{Binless Algorithm Python Script}
\label{sec:binlessPython}

The main work in the binless algorithm is solving the linear system in \myRef{eq:Algorithm} each iteration. Since we would like to apply the binless algorithm to very large data sets, simply forming the matrix $H^{(k)}$ and storing it will require vast amounts of time and memory. Therefore, we need to solve this linear system using an iterative matrix-free method. Such methods do not require the full matrix $H^{(k)}$, but instead just need a method for applying $H^{(k)}$ to a vector (i.e., given a vector $v$, the method returns $H^{(k)}v$).

Our \verb=binless= Python module is a collection of simple functions which, given a data set $x$, provide the discretized calculus operators described in \myRef{sec:binlessImplementation} as either sparse matrices or matrix-free methods. The derivative (and related) operators are very sparse matrices, so the functions which provide them simply return them as NumPy~\cite{SciPy} linked-list sparse matrices. On the other hand, the anti-derivative operators are not sparse, but we can represent them using efficient matrix-free methods, such as using cummulative summation for the left-most matrix in the description of $A$ in \myRef{eq:A}.

We found the iterative solver LGMRES (loose generalized minimum residual algorithm)~\cite{Baker:2005} to be particularly effective for solving the linear system in \myRef{eq:Algorithm}. However, we also found that it was necessary to use a well-chosen preconditioner $P$. A good preconditioner $P$ will approximate the matrix $H^{(k)}$ but will also be easy to invert. Instead of solving the linear system
$H^{(k)} s^{(k)} = -g[u^{(k)}]$, we solve the equivalent system:
\[
	\left(H^{(k)}P^{-1}\right) y = -g[u^{(k)}], \qquad s^{(k)} = P^{-1} y.
\]
Recall that $H^{(k)} = A\T A + \alpha L(u^{(k)})$. We found that the following preconditioner was very effective:
\[
	P = \Diag(\diag(A\T A)) + \alpha L(u^{(k)}),
\]
where $\diag(M)$ returns the diagonal of a square matrix $M$ and $\Diag(v)$ returns a diagonal matrix with the vector $v$ along its diagonal. Using LGMRES with this preconditioner makes the binless algorithm very fast. This allows us to run it multiple times using the background-only data to find the best-fit smoothing parameter $\alpha$ within a short amount of time.

The default run of our \verb=runbinless.py= script generates multiple binless OPDFs as follows:
\begin{enumerate}
  \item Loop over the sample signal plus background UPDFs defined in \\ \verb=utilities/sampleFunctions.py=.
  \item Generate (or load) $N = 1000$ data points for each. Background-only data is also generated.
  \item Iteratively run the binless algorithm on the background data and compare each result to the background-only UPDF to find the best-fit smoothing parameter, $\alpha$.
  \item Using the best-fit $\alpha$, run the binless algorithm on the full signal plus background data.
  \item Save (if necessary) the UPDF data and create the binless plots.
\end{enumerate}

\begin{figure}[t!]
  \centerline{\includegraphics[width=0.8\textwidth]{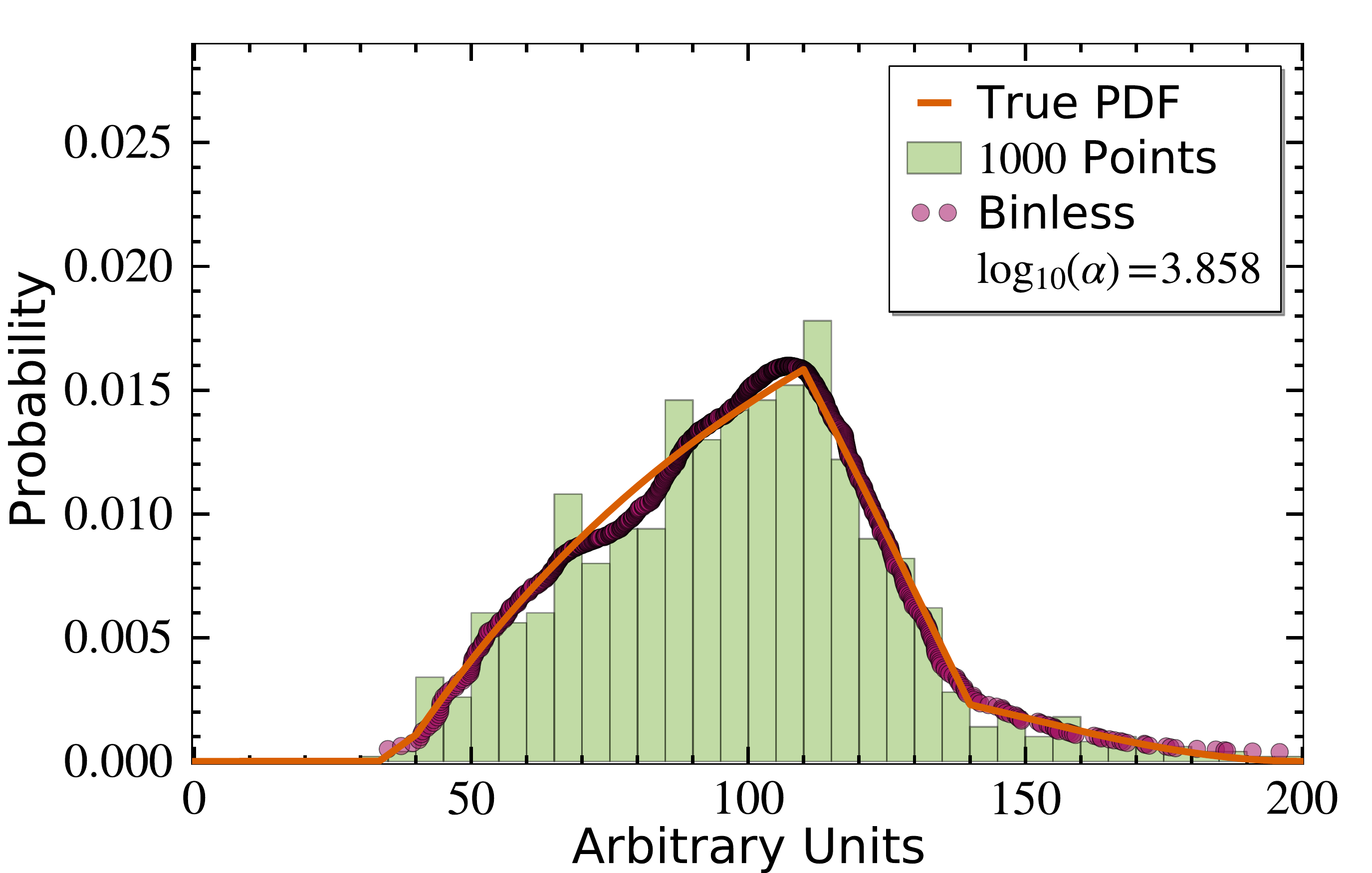}}
  \caption[A comparison between the OPDFs and UPDF for the ``easy endpoint.'' The easy endpoint UPDF is shown as an orange (gray) line. The regular histgoram is filled with 1000 data points, and is shown with green (light gray) filled bars. The binless OPDF is shown as 1000 very densly packed circular points. The best-fit smoothing parameter is given within the legend.]
  {A comparison between the OPDFs and UPDF for the ``easy endpoint'' (\myRef{appx:EasyEndpoint}). The easy endpoint UPDF is shown as an orange (gray) line. The regular histgoram is filled with 1000 data points, and is shown with green (light gray) filled bars. The binless OPDF is shown as 1000 very densly packed circular points. The best-fit smoothing parameter $\alpha$ is given within the legend.}
  \label{fig:binlessEasyEndpoint}
\end{figure}

\begin{figure}[t!]
  \centerline{\includegraphics[width=0.8\textwidth]{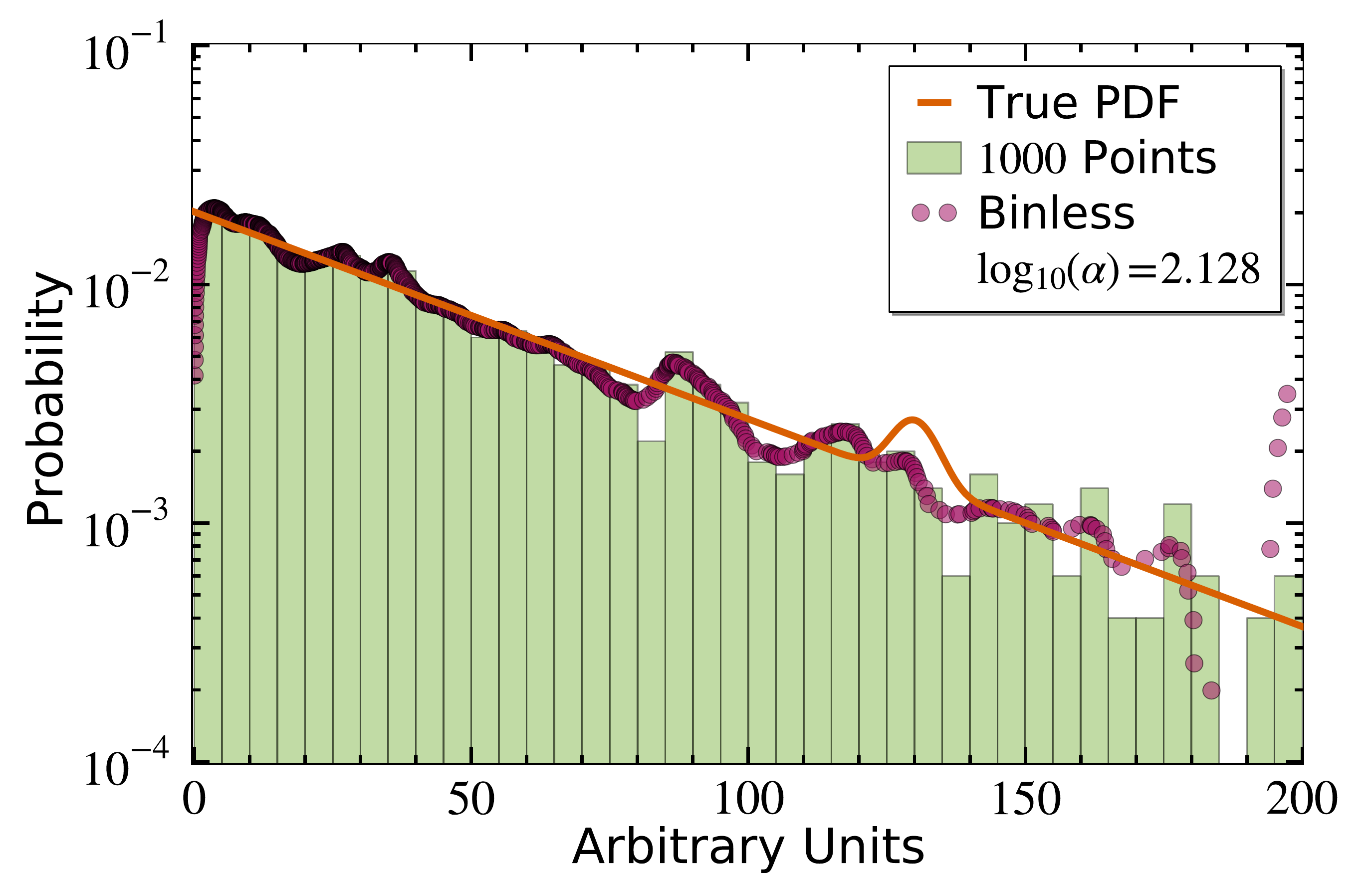}}
  \caption[A comparison between the OPDFs and UPDF for ``the line.'' The line UPDF is shown as an orange (gray) line. The regular histgoram is filled with 1000 data points, and is shown with green (light gray) filled bars. The binless OPDF is shown as 1000 very densly packed circular points. The points near $x=0$ and $x=200$ look particularly bad due to the periodic boundary conditioned derivative operators. The best-fit smoothing parameter is given within the legend.]
  {A comparison between the OPDFs and UPDF for ``the line'' (\myRef{appx:TheLine}).The line UPDF is shown as an orange (gray) line. The regular histgoram is filled with 1000 data points, and is shown with green (light gray) filled bars. The binless OPDF is shown as 1000 very densly packed circular points. The points near $x=0$ and $x=200$ look particularly bad due to the periodic boundary conditioned derivative operators. The best-fit smoothing parameter $\alpha$ is given within the legend.} 
  \label{fig:binlessTheLine}
\end{figure}

We present the results of this process for two sample functions here. The results for the UPDFs described in~\myRef{appx:EasyEndpoint} and~\myRef{appx:TheLine} are shown in~\myRef{fig:binlessEasyEndpoint} and~\myRef{fig:binlessTheLine}, respectively. The former shows an excellent agreement between the binless result and the UPDF. The latter shows the limitation of using periodic boundary conditions within the binless algorithm.

\section{The Binfull Algorithm}
\label{sec:binfull}

The binfull algorithm utilizes a Monte Carlo generator which is based upon the OCDF and can effectively regenerate the original data. To start, we construct the OCDF as $z(x)$ from \myRef{eq:OCDF}. To regenerate a data value, a number, $r$, between zero and one is pulled from a pseudo-random number generator. We multiply by the total number of data values, $N$, and then find the smallest $z(x_i)$ value of the OCDF which is greater than or equal to $rN$. The result is $x_i$.

Like the naive numerical derivative of the OCDF, there is not much use to this Monte Carlo method without data smoothing. A simple example of applying a smoothing function is to add a random deviate, $\epsilon(\sigma)$, which is pulled from a Gaussian distribution of width $\sigma$. Thus, instead of filling our binfull histogram with $x_i$, we fill it with $x_i + \epsilon(\sigma)$. We can run this smoothed Monte Carlo generator an arbitrarily large number of times. The more points it generates, the smaller we can make the bins in our binfull histogram.

\subsection{Binfull Algorithm Python Script}
\label{sec:binfullPython}

The \verb=binfull= module contains utility functions, classes representing different smoothing functions, and a class to contain the binfull histogram resulting from this algorithm. Like a regular histogram, the binfull histogram stores data as a set of bins and bin contents. (Storing all of the raw binfull data turned out to be a memory disaster.)

The default run for \verb=runbinless.py= is as follows:
\begin{enumerate}
  \item Loop over the sample signal plus background UPDFs defined in \\ \verb=utilities/sampleFunctions.py=.
  \item Generate (or load) $N = 1000$ data points for each. Background-only data is also generated.
  \item If binfull data already exists, load it and skip the next two steps.
  \item Iteratively run the binfull algorithm on the background data and compare each result to the background-only UPDF to find the best-fit smoothing parameter, $\sigma$, for each smoothing function.
  \item Using the best-fit $\sigma$ for each smoothing function, run the binless algorithm on the full signal plus background data.
  \item Save (if necessary) the UPDF and binfull data, and create the binfull plots.
\end{enumerate}

\begin{figure}[t!]
  \centerline{\includegraphics[width=0.8\textwidth]{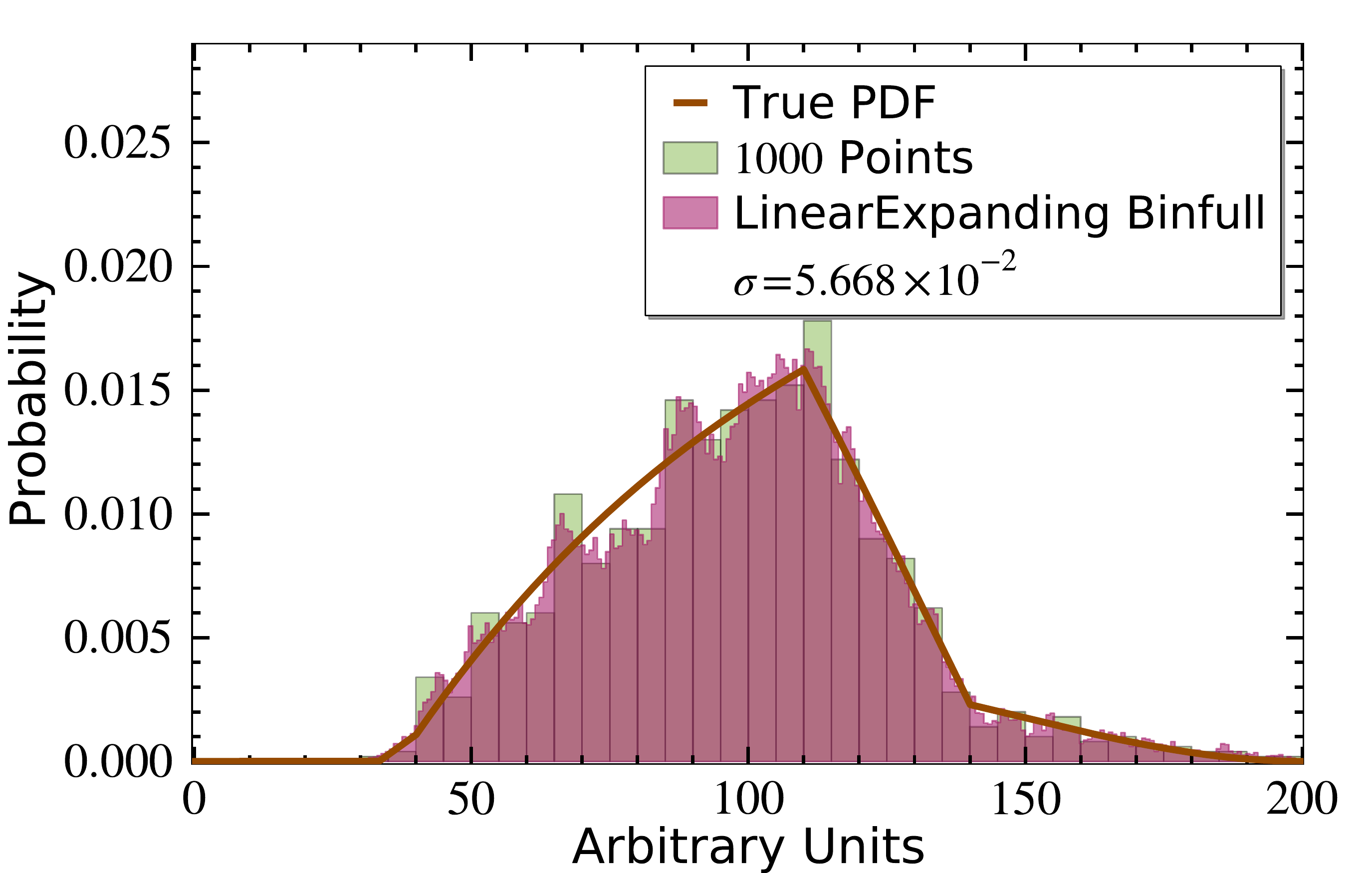}}
  \caption[A comparison between the OPDFs and UPDF for the ``easy endpoint.'' The easy endpoint UPDF is shown as an brown (dark gray) line. The regular histgoram is filled with 1000 data points, and is shown with green (light gray) filled bars. The binless histogram is shown as a purple (gray) filled region. The best-fit smoothing parameter is given within the legend.]
  {A comparison between the OPDFs and UPDF for the ``easy endpoint'' (\myRef{appx:EasyEndpoint}). The easy endpoint UPDF is shown as an brown (dark gray) line. The regular histgoram is filled with 1000 data points, and is shown with green (light gray) filled bars. The binless histogram is shown as a purple (gray) filled region. The best-fit smoothing parameter $\sigma$ is given within the legend.}
  \label{fig:binfullEasyEndpoint}
\end{figure}

\begin{figure}[t!]
  \centerline{\includegraphics[width=0.8\textwidth]{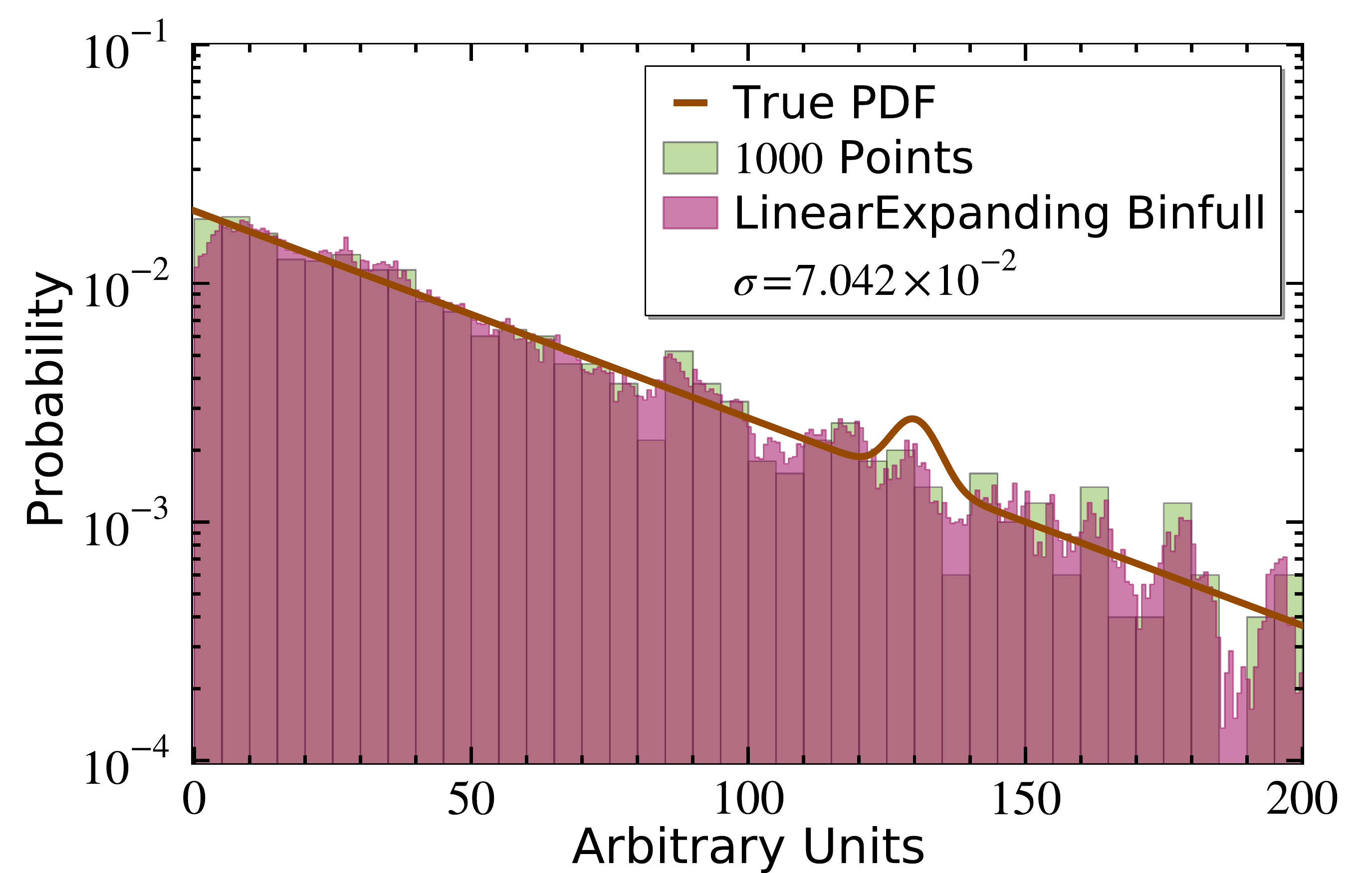}}
  \caption[A comparison between the OPDFs and UPDF for ``the line.'' The easy endpoint UPDF is shown as an brown (dark gray) line. The regular histgoram is filled with 1000 data points, and is shown with green (light gray) filled bars. The binless histogram is shown as a purple (gray) filled region. The best-fit smoothing parameter is given within the legend.]
  {A comparison between the OPDFs and UPDF for ``the line'' (\myRef{appx:TheLine}). The easy endpoint UPDF is shown as an brown (dark gray) line. The regular histgoram is filled with 1000 data points, and is shown with green (light gray) filled bars. The binless histogram is shown as a purple (gray) filled region. The best-fit smoothing parameter $\sigma$ is given within the legend.}
  \label{fig:binfullTheLine}
\end{figure}

We now show the binfull results for the same data sets as we used in~\myRef{sec:binlessPython}. They are shown in~\myRef{fig:binfullEasyEndpoint} and~\myRef{fig:binfullTheLine}, respectively. For these plots, we use a Gaussian smoothing function whose width grows linearly every time the same $x_i$ is generated by the Monte Carlo. Thus, as each point $x_i$ is generated, we fill our binfull histogram with $x_i + \epsilon\bigl(\frac{\sigma \times N_i}{N_{\rm binfull}}\bigr)$, where $N_i$ is the number of $x_i$s given by the Monte Carlo thus far, and $N_{\rm binfull} = 10^5$ is the default number of points generated by the binfull algorithm.

\section{Comparison, Conclusion, and Continuation}
\label{sec:Conclusion}

In this paper we have examined two algorithms for generating representations of OPDFs other than histograms. The binless algorithm determines an OPDF function as the smoothed derivative of the OCDF. The binfull algorithm creates an OPDF histogram which is so full of small bins that it may as well not have bins. Each of these methods has its own strengths and weaknesses.

The binless algorithm is incredibly fast and well suited for larger data sets. The memory requirements of the binless algorithm are small, since for an input data array of $N$ data points, only arrays of length $N$ and $N\times N$ matrices are involved. These matrices are either sparse, or represented as matrix-free methods, each of which returns the operation of the matrix on an array. The result of the binless algorithm is also just an array of length $N$, which is no more memory demanding than the input. Thus, the entire result can be stored for future use.

However, the current implementation of the binless algorithm is only useful for OPDFs which tend towards zero at either end of the domain $\Omega = [0, L]$. This is due to the periodic boundary conditions which we built into the derivative matrices. Additionally, the binless algorithm suffers from a smoothing parameter which is not easy to interpret. It is not at all intuitive as to how $\alpha$ affects the minimization algorithm used to determine the binless OPDF.

The weaknesses of the binless algorithm tend to be the strengths of the binfull algorithm and vice versa. For instance, the smoothing of the binfull algorithm is highly customizable, since one may program their own smoothing function. Because of this, the smoothing parameter used for the binfull algorithm is very intuitive. Also, the binfull algorithm can easily handle OPDFs of any shape, including OPDFs which tend to large values at only one end of the domain.

However, the binfull algorithm is much slower and requires much more memory than the binless algorithm. The default \verb=runBinless.py= script is roughly ten times faster than \verb=runBinfull.py=. The binfull algorithm may be sped up by keeping each and every point of data which it generates within the speedy NumPy arrays. Unfortunately, it then becomes a memory disaster, potentially freezing up the user's computer. Thus, by default, binfull histograms retain only information about their bins and bin contents.

Ultimately, these two methods are quite complementary, each one making up for the weaknesses of the other. Together, they certainly overcome the binning bias inherent in regular histograms. The smoothing parameter they each use is determined blindly, chosen as the value which best reproduces the well-understood background UPDFs.

We even view the shortcomings of these algorithms instead as opportunities for further study. For instance, the determination of smoothing parameters is currently dependent upon the assumption that the background UPDF is known. A data driven method to determine the smoothing parameter or smoothing function would be more ideal. Also, it would be nice to reformulate the problem or the code in order to overcome some of the individual weaknesses of each algorithm, such as flexibility, speed, or memory issues. 

Lastly, it is very important to understand the proper way to statisticlly interpret the results of these algorithms. This is well understood for regular histograms, and crucial for understanding the physics of the UPDF. We are eager to pursue all of these goals in future studies.

\section*{Acknowledgements}
Abram Krislock would like to thank Teruki Kamon and Jan Conrad for useful discussions and follow-up ideas, and Maria Teresa Reynolds for ongoing support and motivation.

\bibliographystyle{hepbib}
\bibliography{biblio}

\appendix
\section{Data Generation}
\label{appx:FakeData}

We generate our test data using a simple Monte Carlo method, much like the one used to generate the binfull OPDF in \myRef{sec:binfull}. For any UPDF, defined on any range of $x$ values, the data generation is as follows. First, an array of $x$ values spanning the range is generated, with a small ($0.01$) step between adjacent values. The UPDF is then evaluated for each of these $x$ values to form an array of $y$ values. Next, the CDF is constructed from the $y$ values using the $A$ operator given by \myRef{eq:A}. Thus, $z_{\rm CDF} = Ay$, where, within $A$, $\Delta_x = 10^{-4}$.

With the CDF generated in this way, we can use it as a Monte Carlo generator of data based upon the UPDF. The generator is the very same as the one described in \myRef{sec:binfull}, except that no smoothing function is applied. This generator can, with infinite statistics, reproduce the shape of the UPDF as given by the $x$ and $y$ arrays.

In this paper, we use the following two phenomenologically inspired UPDFs. For examples of other such UPDFs, please see~\cite{debinning}.

\subsection{Easy Endpoint}
\label{appx:EasyEndpoint}
The ``easy endpoint'' UPDF is inspired by (highly optimistic) endpoint searches of Supersymmetry cascade decays within the context of the Large Hadron Collider or other collider experiment\cite{Hinchliffe:SUSYatLHC}. The easy endpoint UPDF is constructed as a cubic background piece,
\begin{equation}
\label{eq:easyEndpointBG}
  y_{\rm BG} = p_0 (x - p_1)(x - p_2)^2,
\end{equation}
plus a triangular shaped signal piece,
\begin{equation}
\label{eq:easyEndpointSignal}
  y_{\rm signal} = \left\{\begin{array}{cc}
      p_4 x - p_5, & {\rm if}\ x < p_3 \\
      m^\dagger (x - p_6), & {\rm if}\ x \ge p_3
    \end{array} \right.,
\end{equation}
where $p_i$ denote the seven parameters of the function, and $m^\dagger = -\frac{p_3 p_4 - p_5}{p_6 - p_3}$. These pieces are added together to form one function, but only if they are each positive. Thus, the overall easy endpoint UPDF can be written as
\begin{equation}
\label{eq:easyEndpoint}
  y = y_{\rm signal} H(y_{\rm signal}) + y_{\rm BG} H(y_{\rm BG}),
\end{equation}
where $H(y)$ is the Heaviside step function\cite{Heaviside}. The set of seven parameters we use for the easy endpoint are
\begin{equation}
\label{eq:easyEndpointParameters}
  p = (0.00006,\ 40.,\ 200.,\ 110.,\ 1.5,\ 50.,\ 140.).
\end{equation}
We would like to emphasize that the parameter of interest is the location of the ``endpoint,'' which is the maximum $x$-value where the signal piece meets the background. This occurs at $x=p_6=140$.

\subsection{The Line}
\label{appx:TheLine}
The second UPDF we use in this paper is ``the line,'' which is inspired both by the discovery of the Higgs boson at the LHC~\cite{ATLAS:Higgs,CMS:Higgs} and the recent apparent gamma ray line signal within the Fermi-LAT data~\cite{Weniger:Line}. The line UPDF is constructed as an exponential background plus a Gaussian peak signal:
\begin{equation}
\label{eq:theLine}
  y = p_0 \exp(-p_1 x) + p_2 \exp\left(\frac{-(x - p_3)^2}{2 p_4^2}\right).
\end{equation}
The set of parameters we use for the line are
\begin{equation}
\label{eq:theLineParameters}
  p = (1000.,\ 0.02,\ 60.,\ 130.,\ 4.).
\end{equation}
In this case, the parameter of interest is the peak location of the Gaussian, which is $x=p_3=130$.

\end{document}